\documentclass{emulateapj}
\usepackage{amsfonts}
\usepackage{txfonts}
\begin{document}
\title{Strong lensing probability to test TeVeS}
\author{Da-Ming Chen$^1$ and HongSheng Zhao$^{1,2,}$\footnotemark[3]}
\footnotetext[3]{UK PPARC Advanced
Fellow and Overseas Outstanding Youth Fellow}
\affil{$^1$National Astronomical Observatories, Chinese Academy of
Sciences, Beijing 100012, China}
\affil{$^2$SUPA, University of St Andrews, KY16 9SS, Fife, UK}
\begin{abstract}
We calculate the strong lensing probability as a function of the
image-separation $\Delta\theta$ in TeVeS (tensor-vector-scalar)
cosmology, which is a relativistic version of MOND (MOdified
Newtonian Dynamics). The lens, often an elliptical galaxy, is
modeled by the Hernquist profile. We assume a flat cosmology with
$\Omega_b=1-\Omega_\Lambda=0.04$ and the simplest interpolating
function $\mu(x)={\rm min}(1,x)$. For comparison, we recalculated
the probabilities for lenses by  Singular Isothermal Sphere (SIS) galaxy halos
in LCDM with Schechter-fit velocity function. The amplification bias is
calculated based on the
magnification of the second bright image rather than the total of the two
brighter images.  Our calculations show that the Hernquist
model predicts insufficient but acceptable probabilities in flat
TeVeS cosmology compared with the results of the well defined
combined sample of Cosmic Lens All-Sky Survey (CLASS) and Jodrell
Bank/Very Large Array Astrometric Survey (JVAS); at the same time, it
predicts higher probabilities than SIS model in LCDM at small image separations.
\end{abstract}


\section{Introduction}
Since Bekenstein proposed the relativistic, modified Newtonian
dynamics (MOND) theory, named  tensor-vector-scalar
\citep[TeVeS;][]{bekenstein04}, it has become possible to
investigate the MOND phenomena in the cosmological sense. In
particular, after determining the geometry and background evolution
of the Universe, and calculating the deflection of light due to a
weak gravitational field, one can test TeVeS and thus MOND with
gravitational lensing \citep{chiu,zhao06a,angus}. Before TeVeS,
strong gravitational lensing in the MOND regime could only be
manipulated by extrapolating non-relativistic dynamics
\citep{qin,mortlock}, in which the deflection angle is only half the
value in TeVeS \citep{zhaoqin06}.

Needless to say, comparing the predicted results of gravitational
lensing with observations is of key importance in testing TeVeS.
  \citet{zhao06a} first examined the consistency of the
strong lensing predictions in the TeVeS regime for galaxy lenses in
the CfA-Arizona Space Telescope Lens Survey (CASTLES). In this {\it
Letter}, we investigate the  statistics of strong lensing in the
TeVeS regime, and compare the predicted lensing probabilities to the
well defined sample of CLASS/JVAS survey. We adopt the mass function
of the stellar component of galaxies \citep{panters}. As a first
approximation, we do not consider galaxy cluster lenses; the lenses
in the well defined sample in CLASS/JVAS are believed to be produced
by galaxies rather than galaxy cllusters, although a cluster lens, SDSSJ1004,
was discovered in 
Sloan Digital Sky Survey (SDSS) \citep{inada03,oguri04a}. We
consider the simplest MOND interpolating function
$\mu(x)$ and use the Hernquist profile \citep{hernquist} to model
the galaxy lenses. It is now established that, in standard cosmology
(LCDM), when galaxies are modeled by a Singular Isothermal Sphere
(SIS) and galaxy clusters are modeled by a Navarro-Frenk-White (NFW)
profile, the predicted strong lensing probabilities can match the
results of CLASS/JVAS quite well
\citep[e.g.,][]{chae03,chena,chenb,chenc,chend,li02,mitchell05,
oguri02,oguri03b,oguri04,peng06,sarbu01,
wj04,zhang2004,zhang05}. For comparison, we
recalculate the lensing
probabilities predicted by the SIS modeled galaxy lenses in LCDM 
cosmology with the velocity function.
Note that, in LCDM, baryon infall effect
\citep[e.g.,][]{kochanek01,keeton01} has been well described by SIS model for
galaxies \citep{rusin05,koopmans06}, at least statistically;
furthermore, the effects of substructures \citep{oguri06} are also considered
since we use the velocity function to account for the number density of lensing
galaxies. Throughout this {\it Letter}, we assume the source QSOs have a
redshift of $z_s=1.27$.

\section{TeVeS cosmology and deflection angle} 
As in
\citet{bekenstein04} and \citet{zhao06a} we adopt the
Friedmann-Robertson-Walker (FRW) metric in TeVeS, i.e.,
$d\tau^2=-c^2dt^2+a(t)^2[d\chi^2+f^2(\chi)(d\theta^2+\sin^2\theta
d\psi^2)]$ all in physical coordinates, where $c$ is the speed of
light and $f(\chi)=\chi$ for a flat universe.
The proper distance
from the observer to an object at redshift $z$ is
$D^p(z)=c\int_0^z[(1+z)H(z)]^{-1}dz$, where the Hubble parameter at
redshift $z$ is $H(z) \equiv \dot{a}/a \approx H_0\sqrt{\Omega_b(1+z)^3
+\Omega_\Lambda}$, where $\Omega_b$ and $\Omega_\Lambda$ are
the constant density parameter for baryon and dark
energy, respectively, and we set
the contribution from the
scalar field to be zero by approximation \citep{bekenstein04}.  The angular
diameter distance from an object at
redshift $z_1$ to an object at redshift $z_2$, is
$D(z_1,z_2)=[c/(1+z_2)]\int_{z_1}^{z_2}H_0dz/H(z)$ for flat universe. We
assume  
cosmologies with the baryon density $\Omega_b=0.04$
and a Hubble parameter $h=0.73$.

In TeVeS, the lensing equation has the same form as in general relativity (GR),
 and for a spherically symmetric density profile \citep{zhao06a}
\begin{equation}
\beta=\theta-\frac{D_{LS}}{D_{S}}\alpha, \, \, \, \,
\alpha(b)=\int_0^{\infty}\frac{4b}{c^2r}
\frac{d\Phi(r)}{dr}dl,
\end{equation}
where $\beta$, $\theta=b/D_{L}$ and $\alpha(\theta)$ are the source
position angle, image position angle and deflection angle,
respectively; $b$ is the impact parameter; $D_L$, $D_S$ and $D_{LS}$
are the angular diameter distances from the observer to the lens, to
the source and from the lens to the source, respectively; $g(r)=d\Phi(r)/dr$ is
the actual gravitational
acceleration, $\Phi(r)$
is the spherical gravitational potential of the lensing
galaxy and $l$ is the light path. It is well known that the
stellar component of an elliptical galaxy can be well modeled by a
Hernquist profile $\rho(r)=\frac{Mr_h}{2\pi r(r+r_h)^3}$, with the
mass interior to $r$ as $M(r)=\frac{r^2M}{(r+r_h)^2}$, where
$M=\int_0^{\infty}4\pi r^2\rho(r)dr$ is the total mass and $r_h$ is
the scale length. The corresponding Newtonian acceleration is
$g_N(r)=GM(r)/r^2=GM/(r+r_h)^2$. According to MOND
\citep{milgrom83,sanders02,sanders06}, $g(r)\mu(g(r)/a_0)=g_N(r)$.
We choose the simplest interpolating function
$\mu(x)$ with $\mu(x)=x$ for $x<1$ and $\mu(x)=1$ for $x>1$. Thus, the
deflection angle is
\begin{equation}
\alpha(b)=\cases{\int_0^{\sqrt{r_0^2-b^2}}\frac{4GM}{c^2}\frac{bdl}{r(r+r_h)^2}
+\int_{\sqrt{r_0^2-b^2}}^{\infty}\frac{4v_0^2}{c^2}\frac{bdl}{r(r+r_h)},  &for
$b<r_0$, \cr \int_{0}^{\infty}\frac{4v_0^2}{c^2}\frac{bdl}{r(r+r_h)},
&for $b>r_0$, \cr}
\end{equation}
where $r_0$ and $v_0$ are defined by
$GM/(r_0+r_h)^2=v_0^2/(r_0+r_h)=a_0=1.2\times 10^{-8}$cm$s^{-2}$, so that $r_0$
is a transition
radius from the Newtonian to the Mondian regime, $v_0$ is the flat
part of the circular velocity (i.e., the circular velocity in the Mondian
regime). The above deflection angle has
an analytical but cumbersome expression \citep{zhao06a}, so we
calculate it numerically.

\begin{figure}
\epsscale{.80} \plotone{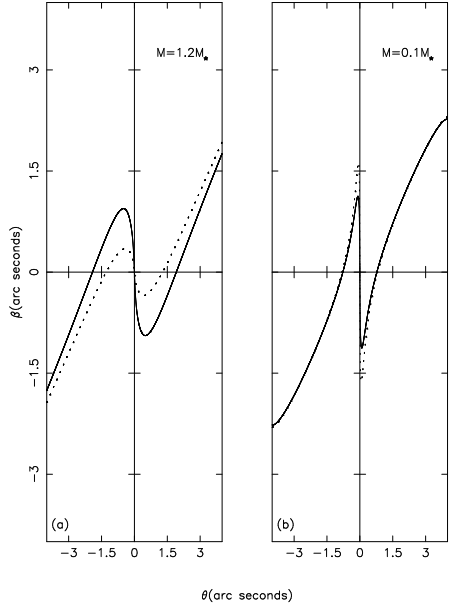} \caption{Source-image relation in a flat
TeVeS cosmology with $z_l=0.05$ and $z_s=1.27$ for a Hernquist lens galaxy with
a mass $M=1.2M_\star$ (panel a) and $M=0.1M_\star$ (panel b).  Note that eq.
(3) predicts a smaller $r_h/r_0$ hence a more effective lens for the $p=0.35$ 
model (solid line) than the $p=0$ model (dashed line) when $M>0.287M_\star$
(as in panel a) and vice versa when $M<0.287M_\star$ (as in panel b).
\label{figlq}}
\end{figure}

We also need a relationship between the scale length $r_h$ and the mass $M$,
which could be determined by observational data. Firstly,
the scale length is related to the effective (or half-light) radius
$R_e$ of a luminous galaxy by $r_h=R_e/1.8$ \citep{hernquist}.  It
has long been recognized that there exists a correlation between
$R_e$ and the mean surface brightness $\langle I_e\rangle$ interior
to $R_e$ \citep{djorgovski}: $R_e\propto\langle I\rangle_e^{-0.83\pm
0.08}$ . Since the luminosity interior to $R_e$ (half-light) is
$L_e=L/2=\pi\langle I\rangle_e R_e^2$, one immediately finds
$R_e\propto L^{1.26}$. Secondly, we need to know the mass-to-light
ratio $\Upsilon=M/L\propto L^{p}$ for elliptical galaxies. The
observed data gives $p=0.35$ \citep{van}; according to MOND,
however, we should find $p\approx 0$ \citep{sanders06}.
In any case we have $L\propto M^{1/(1+p)}$. Therefore, the
scale length should be related to the stellar mass of a galaxy by
$r_h\propto M^{1.26/(1+p)}$. In our actual calculations, we
need to know $r_0/r_h$. Since $(r_0+r_h)\propto M^{1/2}$, we have
$\frac{r_0+r_h}{r_h}=AM^{-p'}$, where $p'={-0.5+1.26/(1+p)}$, and the
coefficient $A$ should
be further determined by observational data. Without a well defined sample at
our disposal, we use the galaxy lenses which have an
observed effective radius $R_e$ (and thus $r_h$) in the CASTLES
survey \citep{munoz}, which are listed in table 2 of
\citet{zhao06a}. The fitted formulae are
\begin{equation}
\frac{r_0}{r_h}+1=\cases{f_1(M)=16.24\times(\frac{M}{0.287M_{\star}})^{-0.43}, &
for $p=0.35$, \cr
f_2(M)=16.24\times(\frac{M}{0.287M_{\star}})^{-0.76}, & for $p=0$}
\label{r0rh}
\end{equation}
where $M_\star=7.64\times 10^{10}h^{-2}M_\sun$ is the characteristic
mass of galaxies \citep{panters}.

Figure \ref{figlq}
shows us the cases when a lens is located at redshift $z=0.05$ but with
different values of $r_0/r_h$ and mass $M$. Here we allow $\beta$
and $\theta$ to take negative values due to symmetry. Generally, three
images are produced when $\beta<\beta_{cr}$, where $\beta_{cr}$ is
the critical source position determined by $d\beta/d\theta=0$ and
$\theta<0$.  For all plausible range of $p=0-0.35$,
Figure \ref{figlq} shows
us that a smaller scale length results in a larger value of
$\beta_{cr}$, as expected.

\section{Lensing probability}
Usually, lensing cross section defined in
the lens plane with image separations larger than $\Delta\theta$ is
$\sigma(>\Delta\theta)=\pi
D_L^2\beta_{cr}^2\Theta[\Delta\theta(M)-\Delta\theta]$, where
$\Theta(x)$ is the Heaviside step function. This is true only when
$\Delta\theta(M)$ is approximately constant within $\beta_{cr}$, and
the effect of the flux density ratio $q_r$ between the outer two
brighter and fainter images can be ignored. From Figure \ref{figlq}
we see that this is not true, in particular for low mass galaxies. As
usual, we consider the outer two images (the central image
is very faint). For a given $\beta$, the left one (with $\theta<0$) is closer to
the center and is fainter, the right one (with $\theta>0$) is further away from
center and is brighter. On the other hand, when $\beta$ increases, the fainter
image approaches to the center and becomes fainter, and opposite for the
brighter image. so larger $\beta$ corresponds to larger flux density ratio, as
is well known. We thus introduce a source position quantity $\beta_{q_r}$
determined by
\begin{equation}
\left(\frac{\theta(\beta)}{\beta}
\frac{d\theta(\beta)}{d\beta}\right)_{\theta>0}=q_r
\left|\frac{\theta(\beta)}{\beta}
\frac{d\theta(\beta)}{d\beta}\right|_{\theta_0<\theta<\theta_{cr}},
\label{qr}
\end{equation}
where $\theta_0=\theta(0)<0$, the absolute value of which is the
Einstein radius, and $\theta_{cr}$ is determined by
$d\beta/d\theta=0$ for $\theta<0$. Equation (\ref{qr}) means that
when $\beta_{q_r}<\beta<\beta_{cr}$, the flux density ratio would be
larger than $q_r$, which is the upper
limit of a well defined sample. For example, in the CLASS/JVAS sample,
$q_r\leq 10$.  The flux
density ratio effect is strongest for intermediate redshift and low mass
lensing galaxies, e.g., for $z \sim 0.5$ and $f_1(M)$,
$\beta_{q_r}/\beta_{cr} \sim 0.35$ at $M=0.1M_\star$ and $\beta_{q_r}/\beta_{cr}
\sim 0.15$ at $M=0.01M_\star$. This effect can be ignored when the redshift of
lensing galaxies $z\sim 0$ or $z\sim z_s$.  On the other
hand, we adopt the suggestion that
the amplification bias should be calculated based on the
magnification of the second bright image of the three images
rather than the total of the two brighter images \citep{lopes}. For
the source QSOs having a power-law flux distribution with slope
$\tilde{\gamma}$ ($=2.1$ in the CLASS/JVAS survey), the amplification
bias is $B(\beta)=\tilde{\mu}^{\tilde{\gamma}-1}$ (Oguri, 2002), where
$\tilde{\mu}(\beta)=|\frac{\theta}{\beta}\frac{d\theta}{d\beta}
|_{\theta_0<\theta<\theta_{cr}}$.

We thus write the lensing cross section with image-separation larger
than $\Delta\theta$ and flux density ratio less than $q_r$ and
combined with the amplification bias $B(\beta)$ as
\citep{schneider92,chenc}

\begin{eqnarray}
&&\sigma(>\Delta\theta,<q_r)=  \nonumber \\
&&2\pi
D_L^2\cases{\int_0^{\beta_{q_r}}\beta
\tilde{\mu}^{\tilde{\gamma}-1}(\beta)d\beta, & for
$\Delta\theta\leq\Delta\theta_0$, \cr
\left(\int_0^{\beta_{q_r}}-\int_0^{\beta_{\Delta\theta}}\right)\beta
\tilde{\mu}^{\tilde{\gamma}-1}(\beta)d\beta, & for
$\Delta\theta_0<\Delta\theta\leq\Delta\theta_{q_r}$, \cr 0, & for
$\Delta\theta>\Delta\theta_{q_r}$.}  \label{cross}
\end{eqnarray}

where  $\beta_{\Delta\theta}$ is the source position at which a lens
produces the image separation $\Delta\theta$,
$\Delta\theta_0=\Delta\theta(0)$ is the separation of the two images
which are just on the Einstein ring, and
$\Delta\theta_{q_r}=\Delta\theta(\beta_{q_r})$ is the upper-limit of
the separation above which the flux ratio of the two images will be
greater than $q_{r}$.

Now we can calculate the lensing probability with image separation
larger than $\Delta\theta$ and flux density ratio less than $q_r$,
in TeVeS cosmology, for the source QSOs at mean redshift $z_s=1.27$
lensed by foreground elliptical stellar galaxies by
\citep[e.g.,][]{wu96}
\begin{eqnarray}
&&P(>\Delta\theta,<q_r)= \nonumber \\
&&\int_0^{z_s}\frac{dD^{p}(z)}{dz}dz\int_0^{M_{max}}n(M,
z)(1+z)^3
\sigma(>\Delta\theta,
<q_r)dM, \label{prob}
\end{eqnarray}
where $M_{max}$ is the upper limit of the mass for the lensing
galaxies,  and $n(M,z)$ is the comoving number density of galaxies for which
we use the well fitted mass
function of the stellar component of galaxies in SDSS given by \citet{panters}:
$n(M)dM=n_\star\left(\frac{M}{M_\star}\right)^{\tilde{\alpha}}
\exp\left(-\frac{M}{M_\star}\right)\frac{dM}{M_\star}$,
where $n_\star=(7.8\pm 0.1)\times 10^{-3}h^3\mbox{Mpc}^{-3}$,
$\tilde{\alpha}=-1.159\pm 0.008$ and $M_\star=(7.64\pm 0.09)\times
10^{10}h^{-2}M_\sun$.  The exact value of $M_{\rm max}$ is unimportant 
but we adopt $M_{\rm max}=10M_\star$ so that 
we do not consider the contribution from galaxy clusters; unlike the
galaxies these are dominated by gas of mass $>10^{12}M_\sun$ with a
$\beta$-profile.

\begin{figure}
\epsscale{.80} \plotone{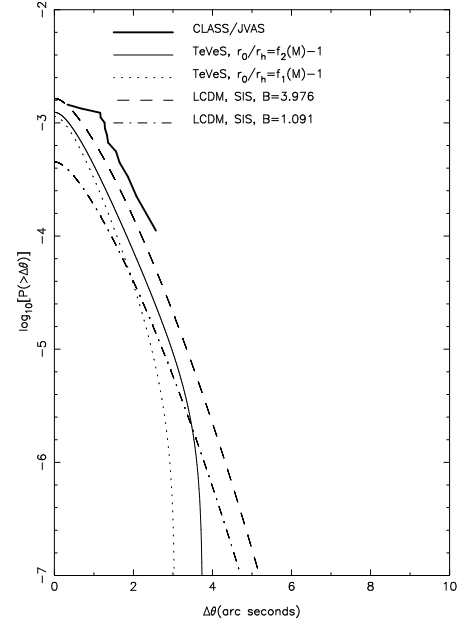} \caption{Predicted lens probability in TeVeS
with an image separation angle $>\Delta\theta$ and the flux
ratio $\le q_r=10$ by Hernquist galaxies. For
comparison, the survey results of CLASS/JVAS
(thick histogram) and the predicted probability for lensing by
SIS halos in LCDM with different amplification bias
($B=3.976$ for dashed line and 1.09 for dash-dotted line) are also shown.
\label{figprob}}
\end{figure}

The numerical results of equation (\ref{prob}) are shown in Figure
\ref{figprob}. The solid line represents the probabilities when
$r_0/r_h=6.3\times(M/M_\star)^{-0.76}-1$
for $M/L=$constant
supported by MOND, and the dotted line represents
$r_0/r_h=9.5\times(M/M_\star)^{-0.43}-1$ for $M/L\propto
L^{0.35}$ from observations. For comparison, the survey results of CLASS/JVAS
\citep{myers,browne03,patnaik92,king99} and the predicted probability for galaxy
lensing by SIS  profiles in LCDM are also shown.  The observational probability
$P_{\mathrm{obs}}(>\Delta\theta)$ \citep{chenb,chenc,chene} is plotted as a
thick histogram in Figure \ref{figprob}. We recalculate the lensing probability
with
image separation larger than $\Delta\theta$ and flux density ratio less than
$q_r$, in flat LCDM cosmology ($\Omega_m=0.3$ and $\Omega_\Lambda=0.7$), for the
source QSOs at mean redshift $z_s=1.27$
lensed by foreground SIS modeled galaxy halos \citep{chae02,ma03,mitchell05}:
$P_{SIS}(>\Delta\theta,<q_r)=\int_0^{z_s}dz\frac{dD^p(z)}{dz}\int_
{v_{\Delta\theta}}^{\infty}dv\bar{n}(v,z)\sigma_{sis}(v,z)B$,
where
$\bar{n}(v,z)dv=n_\star(1+z)^3(\frac{v}{v_\star})^{\tilde{\alpha}}\exp[-(\frac{v
} { v _\star})^{\tilde{\beta}}]\tilde{\beta}\frac{dv}{v_\star}$ is the number
density of galaxy halos at redshift $z$ with velocity dispersion between $v$
and $v+dv$ \citep{mitchell05},
$\sigma_{SIS}(v,z)=16\pi^3(\frac{v}{c})^4(\frac{D_{LS}D_{L}}{D_S})^2$ is the
lensing cross section, $v_{\Delta\theta}=4.4\times
10^{-4}(\frac{c}{v_\star})\sqrt{\frac{D_S\Delta\theta^{''}}{D_{LS}}}$ is the
minimum velocity for lenses to produce image separation $\ge\Delta\theta^{''}$
and $B$ is the amplification bias. We adopt
$(n_\star,v_\star,\tilde{\alpha},\tilde{\beta})=(0.0064h^3\mbox{Mpc}^{-3},
198\mbox{kms}^{-1},-1.0,4.0)$ for early-type galaxies from \citet{chae02}.
According to equation (\ref{qr}) and equation (\ref{cross}), for SIS model and
CLASS sample,
$B=2\int_0^{\beta_{q_r}}\beta\tilde{\mu}(\beta)^{\tilde{\gamma}-1}d\beta$, with
$\beta_{q_r}=9/11$ and $\tilde{\gamma}=2.1$ \citep[see also][]{mitchell05}.
Therefore, $B=1.091$ based on the magnification of the fainter image
(dash-dotted line in Figure \ref{figprob}) and $B=3.976$ based on the total
magnification of two images (dashed line).

\section{Discussion}
It has been held that, in the MOND regime, the
effect of lensing is inefficient, in particular, that strong lensing
never occurs \citep[e.g., ][]{scarpa}. Our calculations shown in
Figure \ref{figprob} indicate, however, that this is not true.
Although the Hernquist model predicts insufficient lensing
probabilities in a flat TeVeS cosmology compared with the result of CLASS/JVAS, 
the result is acceptable
considering, at least, that the lensing galaxy can be modeled by
steeper slopes and more efficient MOND $\mu$-functions. 

Our results argue that TeVeS (and thus MOND) generates lenses with higher
efficiency than CDM if the latter is modelled by SIS profile and in both
cases the amplification bias  is calculated based on the magnification
of the second bright image (for SIS, the fainter image).  Usually,
$B$ is calculated based on the total magnification of the two images cosidered.
Because we introduced a cutoff $\beta_{q_r}$ due to the flux ratio
$q_r$, the total magnification is $2\sim q_r+1$ times larger than that of the
second bright image (depending on $\beta$ when $\beta\le\beta_{q_r}$), which
results in the corresponding value of $B$ about 4 times larger (both for SIS in
LCDM and Hernquist in TeVeS). As shown in Figure \ref{figprob}, the lensing
probabilities for SIS halos in
LCDM with $B=3.976$ (total) matches the results of CLASS/JVAS quite well (dashed
line). Similarly, if we apply the total magnification to $B$ for
Mondian Hernquist model, the final lensing probabilities would be overpredicted
compared with the results of CLASS/JVAS.
Note that an SIS profile is more concentrated in mass than a
Hernquist profile, so if both profiles are applied in the same
regime (LCDM or TeVeS), the SIS profile would be more effective at
lensing than Hernquist. Therefore, the fact that the probabilities
for SIS model in LCDM ($B=1.09$, dash-dotted line) are lower than the Hernquist
model in TeVeS
shown in Figure \ref{figprob} implies that MOND demonstrates a
higher lensing efficiency than CDM. This phenomena is, in fact,
not difficult to understand. It is well known that MOND, as an
alternative to dark matter for solving the ``missing mass" problem,
takes effect in the region surrounding the luminous matter with
$r>r_0$, where a CDM halo is assumed to have non-zero density and
its acceleration dominates over luminous matter in LCDM cosmology
\citep{kaplinghat}. The deflection angle $\alpha(b)$ with impact
parameter $b>r_0$ can be calculated using Newtonian CDM gravitation
or Mondian luminous matter gravitation. We know that the
acceleration $g(r)$ in the equation for deflection angle for an SIS
modeled CDM halo is $g(r)\propto r^{-1}$, independent of $b$. So,
the image separation is independent of the source position angle
$\beta$ (when $\beta<\beta_{cr}$), as is well known in the SIS
model. However, for a lensing galaxy (with no dark matter) modeled
by a Hernquist profile, we have $g(r)\propto r^{-1}$  only when
$r>r_0$ (Mondian regime). So the higher probabilities indicate a higher
lensing efficiency between MOND and CDM. 

As a first attempt at investigating strong lensing statistics in the
TeVeS scenario, we have used the simplest interpolating function
$\mu(x)$. The deflection angle is, of course, sensitive to $\mu(x)$
\citep{zhao06b}. The simplest $\mu(x)$ adopted in this {\it Letter}
corresponds to the lowest physical (or ``true") acceleration $g(r)$.
Any other forms of $\mu(x)$ all give stronger
physical accelerations than the simplest one \citep{zhaotian06}.
Furthermore, strong lensing  is very
sensitive to the concentration and the slope near the center of the
density profile of lensing galaxies.  The Hernquist model and 
an NFW model has $\rho(r)\sim r^{-1}$ near the center, both are inefficient in
lensing. If elliptical galaxies were modeled as pressure- supported
Jaffe model, e.g.,  with $\rho(r)\sim \frac{1}{r^{2}(r+a)^2}$ where
$a$ is a core scale length, then the lensing probability would be
increased.  Another important point is that we have assumed a flat
universe with $\Omega_\Lambda=0.96$. However, by fitting to high-z
SN Ia luminosity modulus, \citet{zhao06a} showed that an open
universe is more likely in TeVeS (with $\Omega_\Lambda=0.5$). In
summary, it is promising to constrain TeVeS vs CDM through lensing statistics.

\acknowledgments{We are grateful to the anonymous referee for his 
insightful  comments and helpful suggestions. This work was supported by the
National Natural Science Foundation of China under grant 10233040 and 10673012
to D.-M. C and
10428308 to H. Z.}

\end{document}